\documentclass[%
 reprint,
showpacs,
 amsmath,amssymb,
 aps,
pra,
floatfix,
]{revtex4-1}

\usepackage[colorlinks=true, linkcolor=blue, citecolor=blue, urlcolor=blue, filecolor=blue]{hyperref}		
\usepackage{graphicx}
\usepackage[caption=false]{subfig}

\begin{document}

\preprint{APS/123-QED}

\title{Kinetics of Fluid Demixing in Complex Plasmas: Domain Growth Analysis using Minkowski Tensors}

\author{A. B\"obel}
	\email{alexander.boebel@dlr.de}
\author{C. R\"ath}%
 
\affiliation{%
 Forschungsgruppe Komplexe Plasmen, Deutsches Zentrum f\"ur Luft- und Raumfahrt (DLR), Argelsrieder Feld 1a, 82234 Wessling, Germany\\
}

\date{June 25, 2016}


\renewcommand{\figurename}{FIG.}

\begin{abstract}
A molecular dynamics simulation of the demixing process of a binary complex plasma is analysed and the role of distinct interaction potentials is discussed by using morphological Minkowski tensor analysis of the minority phase domain growth in a demixing simulated binary complex plasma. These Minkowski tensor methods are compared with previous results that utilized a power spectrum method based on the time-dependent average structure factor. It is shown that the Minkowski tensor methods are superior to the previously used power spectrum method in the sense of higher sensitivity to changes in domain size. By analysis of the slope of the temporal evolution of Minkowski tensor measures qualitative differences between the case of particle interaction with a single length scale compared to particle interactions with two different length scales (dominating long range interaction) are revealed. After proper scaling the graphs for the two length scale scenario coincide, pointing towards universal behaviour. The qualitative difference in demixing scenarios is evidenced by distinct demixing behaviour: In the long range dominated cases demixing occurs in two stages. At first neighbouring particles agglomerate then domains start to merge in cascades. However in the case of only one interaction length scale only agglomeration but no merging of domains can be observed. Thus, Minkowski Tensor analysis are likely to become a useful tool for further investigation of this (and other) demixing processes. It is capable to reveal (nonlinear) local topological properties, probing deeper than (linear) global power spectrum analysis, however still providing easily interpretable results founded on a solid mathematical framework.
\end{abstract}

\maketitle

\section{Introduction \label{sec1}}
When a binary fluid is forced into the immiscible state, it starts to dynamically demix until the thermodynamically stable state of two coexisting fluids is reached. This spinodal decomposition is accompanied by a domain growth that is believed to be selfsimilar in time; i.e., the domain morphology is preserved. This implies a single time dependent characteristic length which obeys a power law growth $L(t) \propto t^{\alpha}$ \cite{dmx1_doi:10.1080/00018730110117433, dmx2_1367-2630-7-1-040, dmx3_PhysRevLett.99.205701, dmx_lj_PhysRevE.77.011503, dmx4_0295-5075-93-5-55001,demixing2010_PhysRevLett.105.045001, diff_demix}.

Competing interactions play an important role in the morphology of phase separation. They can lead to various domain patterns as e.g. striped lamellar structures, hexagonal arrays of droplets \cite{dmx5_Seul27011995} and clusters \cite{dmx6}. The dynamical evolution in systems with such interactions is governed by the competition between the long range repulsion causing subdivision of domains and the short range attraction resulting in the growth of interface energy. There has been a great deal of theoretical research of this process carried out in the mean-field framework \cite{dmx7_PhysRevE.49.2225, dmx8_PhysRevB}. However, there are few particle resolved studies of such systems \cite{dmx9_0953-8984-17-45-028}, in particular few studies exist addressing the role of competing interactions \cite{demixing2010_PhysRevLett.105.045001}.

Complex plasmas are composed of a weakly ionized gas and microparticles which are highly charged due to absorption of the ambient electrons and ions \cite{dmx10, dmx11_RevModPhys.81.1353}. Binary complex plasmas contain microparticles of two different sizes and constitute a model system which is well suited for studying the kinetics of fluid demixing at the individual particle level: Properties of pair interactions, such as the interaction range, can be flexibly tuned. Also the dynamics of particles at short time scales is practically undamped due to the low density in the gas in typical complex plasmas \cite{dmx11_RevModPhys.81.1353}.

The prevailing mechanism of interaction between charged microparticles in complex plasmas is electric repulsion \cite{dmx10, dmx11_RevModPhys.81.1353}. At large distances $r$, the electrostatic potential of a particle can be represented in the asymptotic form $\Phi (r)$ \cite{dmx17_0295-5075-85-4-45001}. Theory predicts a rich variety of screening mechanisms operating in complex plasmas \cite{dmx10, dmx12_PhysRevLett.100.225003}. The shape of $\Phi (r)$ can be affected by the plasma absorption on a particle, nonlinearity in plasma particle interactions, ionization loss balance, etc. Recently, it was shown that the plasma production and loss processes can play a crucial role in the long range behavior of $\Phi (r)$ \cite{dmx18} resulting in the emergence of two dominating asymptotes that both have the Yukawa form:
\begin{equation}
\Phi (r) \; = \; \dfrac{1}{r} \left( Z^{*}_{SR} e^{-r/\lambda_{SR}} + Z^{*}_{LR} e^{-r/\lambda_{LR}}\right)
\label{eq-yukawa}
\end{equation}
Here $Z^*$ indicates effective charge, $\lambda$ indicates the length scale for long range (LR) and short range (SR) interactions respectively. 
The screening length ratio is defined as
\begin{equation}
\Lambda \; = \; \lambda_{LR} / \lambda_{SR}
\label{eq-lamda}
\end{equation}
The following discussion will be based on simulations where $\Lambda$ is varied since it was shown that the LR interactions can significantly enhance demixing \cite{demixing2010_PhysRevLett.105.045001}.

To study the possibility of demixing in binary complex plasmas experiments with the PK-3 Plus rf discharge chamber \cite{dmx15_1367-2630-10-3-033036} on board the ISS were performed earlier. After the injection of small particles in a stationary cloud of big ones an apparent phase separation was observed accompanied by the formation of a small particle droplet (supplementary movie S1 of \cite{demixing2010_PhysRevLett.105.045001}). These experiments showed the strong tendency of binary complex plasmas to demix at time scales of seconds.

Since the early 20th century \cite{Minkowski1903} Minkowski functionals are a prominent tool for morphological data analysis. Only recently the hierarchy of Minkowski valuation was extended to tensor valued quantities called Minkowski tensors \cite{mink_alesker}. Minkowski functionals and tensors are sensitive to any $n-$ point correlation function and thus can quickly give new insights to processes beyond the capability of power spectrum methods as demonstrated for the demixing of a binary complex plasma in this paper.

This paper is structured as follows. Section \ref{sec-simulation} describes the simulations on which the Minkowski tensor analysis is performed upon. In Section \ref{sec-mts} methods are presented: Voronoi tesselations (\ref{subsec-voronoi}), Minkowski functionals (\ref{subsec-mf}) and Minkowski tensors (\ref{subsec-mt}) are introduced. Based on this introduction an isotropy mesasure (\ref{subsec-beta}) and a symmetry metric (\ref{subsec-delta}) is derived. Also the previously used power spectrum method \ref{subsec-ps} is reviewed. Section \ref{sec-results} presents the results obtained by Minkwski tensor analysis of the simulation data. The dynamic range (\ref{subsec-dyn}) of the Minkowski measures, differences in demixed domain size (\ref{subsec-diff}) for different measures are discussed and hints of universal behaviour (\ref{subsec-univ}) can be found by analysis of local gradients of the temporal evolution of Minkowski measures. Finally in section \ref{sec-conclusion} conclusions are drawn and a brief outlook on more detailed studies of this process is given.
\section{Simulation \label{sec-simulation}}
In order to investigate details of the particle dynamics accompanying the phase separation in binary complex plasmas and compare with theory molecular dynamics (MD) simulations with the Langevin thermostat were employed previously: The growth of the minority phase (particle species 1) domains was analysed for several combinations of short- and long range interactions using a power spectrum method based on the time dependent average structure factor. \cite{demixing2010_PhysRevLett.105.045001} A binary mixture was composed of, in total, $729 000$ particles (of species 1 and 2) at the off-critical particle composition $x_1 = 0.5$ (equal number of particles of species 1 and 2). The simulations were performed in a cubic box with the dimensions of $27$ mm (corresponds to a mean interparticle distance of $0.3$ mm) and periodic boundary conditions. The particles interacted via the potential (\ref{eq-yukawa}). Following simulation parameters (approximately corresponding to the experiment) were used: The particle mass density 1.5
g/cm$^3$ and diameters $2a_1=3.4$ $\mu$m and $2a_2=9.2$ $\mu$m, the actual charges $Z_1=4000~e$ and
$Z_2=(a_2/a_1)Z_1=10824~e$, the friction coefficients $\zeta_1=250$ s$^{-1}$ and $\zeta_2=(a_1/a_2)\zeta_1=92.4$ s$^{-1}$,
the SR (Debye-H\"uckel) screening length $\lambda_{\rm SR}=150~\mu$m, the mean interparticle distance $\Delta=0.3$~mm, and
the temperature $k_{\rm B}T=0.024$ eV. A standard integration scheme was employed with a time step
$\delta t=0.0025<\zeta_{1}^{-1}<\zeta_{2}^{-1}$~s to solve the equation of motion numerically. Further details can be found in the supplement of  \cite{demixing2010_PhysRevLett.105.045001}.

Here the results of these earlier simulations are analysed in terms of Minkowski tensor methods and compared with previous results obtained by a power spectrum analysis.
\section{Minkowski Tensor Methods \label{sec-mts}}
\subsection{Voronoi Tessellation \label{subsec-voronoi}}
A commonly used method for quantifying the local structure of spheres is by construction of a nearest neighbour network on which quantitative structure metrics are computed (e.g. bond orientational order parameters \cite{voronoi4_PhysRevB.28.784}).

An alternative approach for quantifying local structure is provided by the analysis of the Voronoi diagram. The Voronoi diagram is the partition of space into the same number of convex cells as there are particles in the packing. The Voronoi cell of each particle is the region of space closer to that given particle than to any other particle. For the special case of three dimensional crystal lattices the Voronoi cell is called Wigner-Seitz cell. In the field of granular matter, Voronoi diagrams have been used to determine distributions of local packing fractions \cite{voronoi6_PhysRevLett.96.018002, voronoi7_0295-5075-79-2-24003, voronoi8_PhysRevE.77.021309}, spatial correlations \cite{voronoi9_0295-5075-97-3-34004} and correlations with particle motion \cite{voronoi10_PhysRevLett.101.258001}.

Recently, studies provided insight in the local structure of sphere packings and sphere ensembles by analysing the shape of Voronoi cells, in particular their degree of anisotropy or elongation \cite{voronoi11_0295-5075-90-3-34001, voronoi12_1742-5468-2010-11-P11010, voronoi13_kapfer2012a, voronoi}.

Here the shape of the Voronoi tessellation obtained from particle positions is analysed using Minkowski functional and tensor methods. The boundary particles where discarded, the tessellation included only a center cube with an edge length of $16 mm$.
\subsection{Minkowski Functionals \label{subsec-mf}}
For a body $K$ with a smooth boundary contour $\partial K$ embedded in $D$-dimensional euclidean space the $D+1$ Minkowski functionals are, up to constant factors, defined as:
\begin{equation}
\begin{aligned}
W_0(K) \; &= \; \int_K \, \mathrm{d}^D r
\\
W_{\nu} (K) \; &= \; \int_{\partial K} \, G_{\nu}(r) \, \mathrm{d}^{D-1} r \quad , \quad 1 \leq \nu \leq D
\end{aligned}
\label{eq-mf}
\end{equation}
$G_{\nu}(r)$ are the elementary symmetric polynomials of the local principal curvatures as defined in differential geometry.

The Minkowski tensor methods described in the following are also implemented for two dimensional systems and thus can be applied to e.g. experimental observations of slices of a three dimensional system. In three dimensinal euclidean space the Minkowski functionals, up to constant factors, are $W_0(K)$ (volume), $W_1(K)$ (area),$W_2(K)$ (integrated mean curvature) and $W_3(K)$ (euler characteristic):\begin{equation}
\begin{aligned}
W_0(K) &= \int_K \mathrm{d}^3 r 
\\
W_1(K) &= \int_{\partial K} \mathrm{d}^2 r 
\\
W_2(K) &= \int_{\partial K} \kappa_1 + \kappa_2 \: \mathrm{d}^2 r 
\\
W_3(K) &=\int_{\partial K} \kappa_1  \kappa_2 \: \mathrm{d}^2 r 
\end{aligned}
\label{eq-mf3d}
\end{equation}

Minkowski functionals are motion invariant, additive and (conditionally) continuous. They form a complete family of morphological measures. Or vice versa: Any motion invariant, (conditionally) continuous and additive functional is a superposition of the (countably many) Minkowski functionals. They are nonlinear measures sensitive to higher order correlations. Known applications are e.g. curvature energy of membranes \cite{mink_app1}, order parameter in Turing patterns \cite{mink_app2_Mecke_turing}, density functional theory for fluids (as hard balls or ellipsoids) \cite{mink_app3_density_func, mink_app_mecke}, testing point distributions (find clusters, filaments, underlying pointprocess) or searching for non-Gaussian signatures in the CMB \cite{mink_app_kerscher}.

\begin{figure}[!tbp]
	\captionsetup[subfigure]{position=top,singlelinecheck=off,labelfont=bf,textfont=normalfont, justification=raggedright, margin=10pt,captionskip=-1pt }
	\hspace*{-0.2cm} 
	\subfloat[\label{sub_vol-1}]{%
	\includegraphics[width=0.24\textwidth]{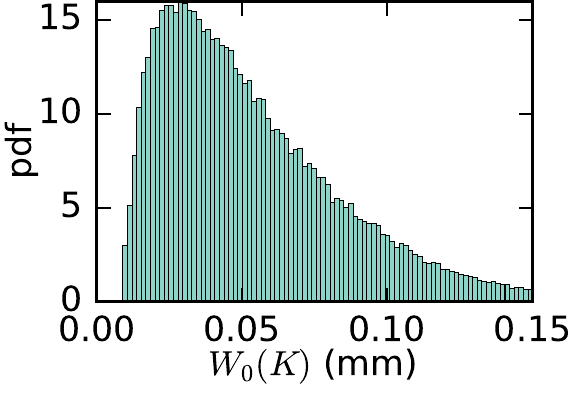}}
	\hspace*{-0.1cm} 
	\subfloat[\label{sub_vol-2}]{%
	\includegraphics[width=0.24\textwidth]{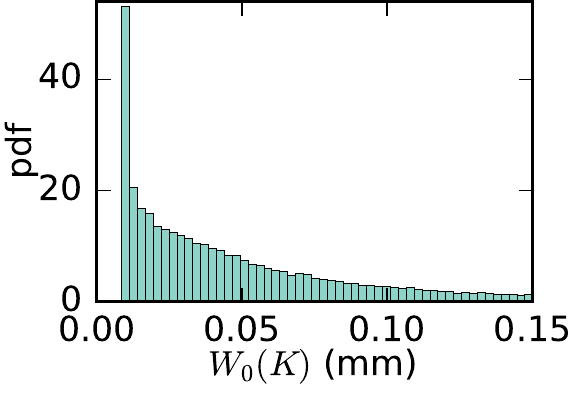}}
	\vspace{-0.5cm} 
	\hspace*{-0.2cm}
	\subfloat[\label{sub_vol-3}]{%
	\includegraphics[width=0.24\textwidth]{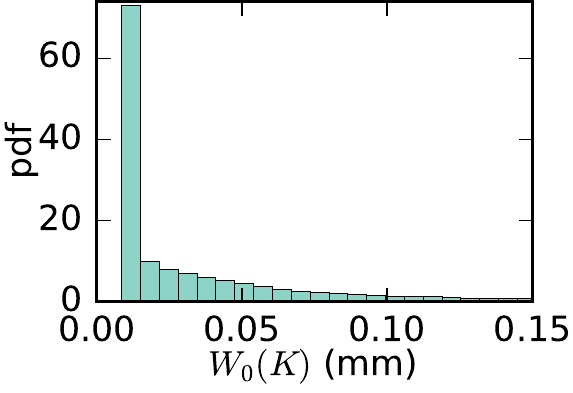}}
	\hspace*{-0.1cm} 
	\subfloat[\label{sub_vol-4}]{%
	\includegraphics[width=0.24\textwidth]{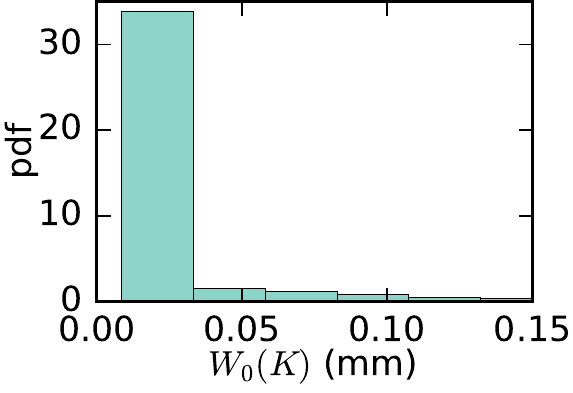}}
\caption{\label{fig-volume_hist} \small Calculating the minority phase (particle species 1) domain size via a Histogram method. The volume functional $W_0(K)$ (\ref{eq-mf3d}) of species 1 particles is plotted for increasing time from the top left to the bottom right panel. (a) $t=0.6 \, s$, (b) $t=0.6 \, s$, (c) $t=0.6 \, s$, (d) $t=0.6 \, s$. The ever increasing low volume peak can be interpreted as ordered domains whereas disordered particles that not yet agglomerated correspond to larger Voronoi regions. These plots are obtained for simulations with screening length ratio $\Lambda = 12$. For other values of the screening length ratio similar plots are obtained.}
\end{figure}

The Minkowski functional analysis carried out in this study is done by calculating the volume functional $W_0(K)$  (\ref{eq-mf3d}) locally for every Voronoi region of the minority phase (particle species 1) and for every time step in the simulation. Then a histogram is calculated for every time step. As illustrated in Fig. \ref{fig-volume_hist} these histograms, as time progresses, separate into two parts: The narrow delta shaped peak corresponding to the smallest volume becomes even more prominent and distinct from the larger volume tail. It is reasonable to assume that the smallest volume peak corresponds to the homogeneous domains of agglomerating particles. To measure the size of these domains the number of Voronoi regions in the smallest volume peak of the histogram are counted. The hereby obtained value will hereinafter be referred to as MT0 measure.

\subsection{Minkowski Tensors \label{subsec-mt}}

In order to account also for directional properties it is natural to abstract the scalar valued Minkowski functionals to tensor valued quantities called Minkowski tensors \citep{rev_MT_turk}:
\begin{equation}
\begin{aligned}
W_0^{a,0}(K) \; &:= \; \int_K \, \mathrm{d}^D r \; \; \mathbf{r}^{\odot a}
\\
W_{\nu}^{a,b} (K) \; &:= \; 1/D  \int_{\partial K}  \mathrm{d}^{D-1} r  \; \; G_{\nu}(r) \; \; \mathbf{r}^{\odot a} \odot \mathbf{n}^{\odot b}
\end{aligned} 
\label{eq-mt}
\end{equation}
Here $\odot$ stands for the symmetric tensor product. Again $G_{\nu}(r)$ are the elementary symmetric polynomials of the local principal curvatures as defined in differential geometry. $a$ counts the number of position vectors $\mathbf{r}$, $b$ counts the number of normal vectors $\mathbf{n}$ in the tensor product. Thus the rank of each tensor is the tupel $(a,b)$.
Their properties are as follows: They are isometry covariant, i.e their behaviour under translation and rotation is given by:
        \begin{equation}
        \begin{aligned}
        W_{\nu}^{a,b} (K + \mathbf{t}) \; &= \; \sum_{i=0}^{a} \, \binom{a}{i} \, \mathbf{t}^i \, W_{\nu}^{a-i,b} (K)
\\
        W_{\nu}^{a,b} ( \hat{O} \, K) \; &= \; \hat{O}_{a+b} \;  W_{\nu}^{a,b} (K)
        \end{aligned}
		\label{eq-mt-trans}
		\end{equation}
They are additive
\begin{equation}
W_{\nu}^{a,b}(K_1 \cup K_2) = W_{\nu}^{a,b}(K_1) + W_{\nu}^{a,b}(K_2) - W_{\nu}^{a,b}(K_1 \cap K_2)
\end{equation}
and they are homogeneous of degree $3+a-\nu$:
\begin{equation}
W_{\nu}^{a,b} ( \lambda K) \; = \; \lambda^{3+a-\nu} \;  W_{\nu}^{a,b} (K)
\end{equation}

Similar to Minkowski functionals the attractiveness of Minkowski tensors is particularly due to a strong completeness theorem by Alesker \cite{mink_alesker}. It is stating that all morphological information that is relevant for additive material properties is represented by the Minkowski tensors. Any motion covariant, conditionally continuous and additive tensor valued functional is a superposition of the (countably many) Minkowski Tensors.
\subsection{MT2 Isotropy Index \label{subsec-beta}}
For a body $K$ and each rank two Minkowski tensor $W_{\nu}^{a,b}(K)$ an isotropy index can be defined as the ratio between the smallest and largest eigenvalue of the $D \times D$-matrix representing each Minkowski tensor: \cite{rev_MT_turk}
\begin{equation}
\beta_{\nu}^{a,b}(K) : = \dfrac{\lambda_{min} \left( W_{\nu}^{a,b}(K) \right) }{\lambda_{max} \left( W_{\nu}^{a,b}(K) \right)}
\label{eq-beta}
\end{equation}
The dimensionless isotropy index is a pure shape measure. It is invariant under isotropic scaling of $K$. For example in two dimensions the isotropy index $ \beta =1 $ is obtained for a circle or a square. For a rectangle one obtains $ \beta = shorter/longer \, edge$. In three dimensions $ \beta = 1 $ is obtained for regular shapes ranging from a cube to a sphere. For a box the value  is $ \beta = shortest/longest \, edge$.
Thus this isotropy index is an isotropy measure only in the sense of elongation.

\begin{figure}[!tbp]
	\captionsetup[subfigure]{position=top,singlelinecheck=off,labelfont=bf,textfont=normalfont, justification=raggedright, margin=10pt,captionskip=-1pt }
	\hspace*{-0.2cm} 
	\subfloat[\label{sub_w120-1}]{%
	\includegraphics[width=0.24\textwidth]{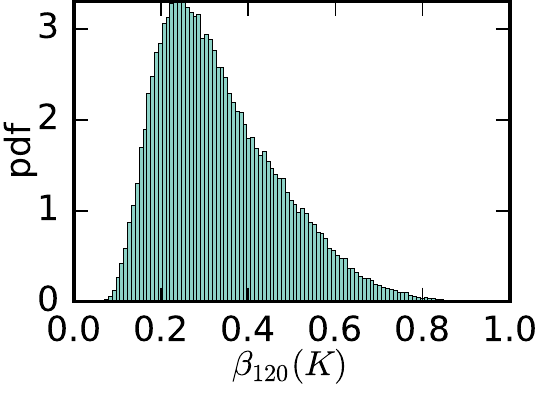}}
	\hspace*{-0.1cm} 
	\subfloat[\label{sub_w120-2}]{%
	\includegraphics[width=0.24\textwidth]{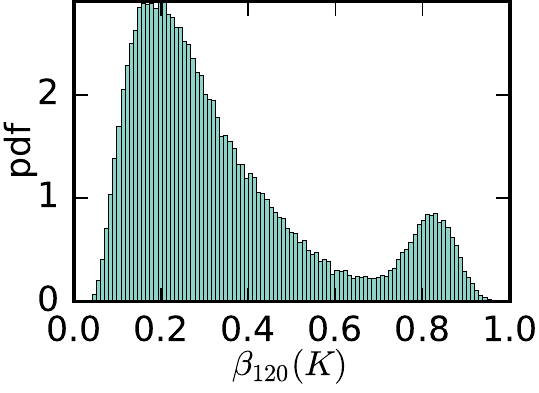}}
	\vspace{-0.5cm} 
	\hspace*{-0.2cm}
	\subfloat[\label{sub_w120-3}]{%
	\includegraphics[width=0.24\textwidth]{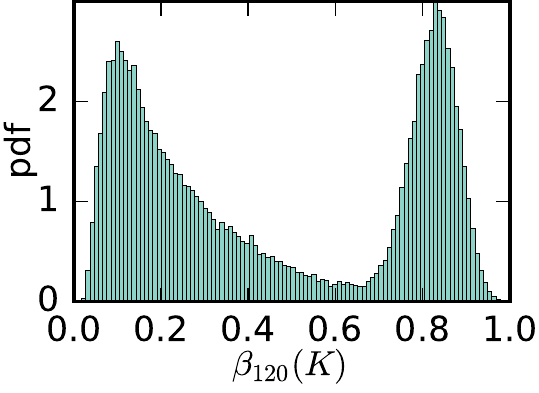}}
	\hspace*{-0.1cm} 
	\subfloat[\label{sub_w120-4}]{%
	\includegraphics[width=0.24\textwidth]{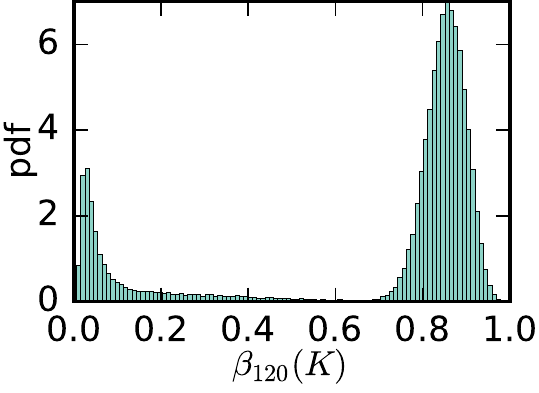}}
\caption{\label{fig-w120_hist} \small Calculating the minority phase (particle species 1) domain size via a Histogram method. The isotropy index $\beta$ (\ref{eq-beta}) of minority phase particles is plotted for increasing time from the top left to the bottom right panel. (a) $t=0.6 \, s$, (b) $t=0.6 \, s$, (c) $t=0.6 \, s$, (d) $t=0.6 \, s$. The separating parts can be interpreted as ordered domains (high $\beta$ values) and disordered particles that not yet agglomerated (low $\beta$ values) into a order domain. These plots are obtained for simulations with screening length ratio $\Lambda = 12$. For other values of the screening length ratio similar plots are obtained.}
\end{figure}

The rank two Minkowski tensor analysis carried out in this study is done by calculating the isotropy index $\beta$ (\ref{eq-beta}) locally for every Voronoi region of the minority phase (particle species 1) and for every time step in the simulation. Then a histogram is calculated for every time step. As illustrated in Fig. \ref{fig-w120_hist} these histograms are composed of two parts: The ordered, isotropic section (high $\beta$ values) is separating from the disordered, anisotropic (small $\beta$ values) section with increasing time steps. It is reasonable to assume that the isotropic Voronoi regions correspond to the homogeneous domains of agglomerating particles. To measure the size of these domains the number of Voronoi regions in the ordered, high $\beta$ section of the histogram are counted. To that purpose, in order to distinguish ordered from disordered sections a cut off value ($\beta_{thresh} = 0.7$) is chosen as the intersection point of the two sections at late time steps when demixing is well progressed. For simplicity this measure will hereinafter be referred to as $MT2$ measure.

\subsection{MT4 Symmetry Metric \label{subsec-delta}}

In order to distinguish between structures of high symmetry, i.e. differentiate between crystalline structures (hcp, fcc, etc.) higher order tensors have to be applied. For rank four and higher, isotropic symmetry is distinct from cubic symmetry. (This is evidenced by the appearance of a second independent shear modulus when transitioning from isotropic to cubic symmetry in the theory of linear elasticity, which is formulated using a rank-four tensor \cite{walpole_elastic_theory}. This method has been used in hard sphere systems to characterise random close packings \cite{jammed_spheres_mt4}.

For brevity only the simplest rank four Minkowski tensor is considered: 
\begin{equation}
W_1^{04} (K) = 1/3 \cdot \int_{\partial K} \mathrm{d}^2 r \; \mathbf{n}(\mathbf{r}) \otimes \mathbf{n}(\mathbf{r}) \otimes \mathbf{n}(\mathbf{r}) \otimes \mathbf{n}(\mathbf{r}).
\end{equation}
Since it is translation invariant and symmetric (i.e. it holds for the components $\left[ W_1^{04} \right]_{ijkl} = \left[ W_1^{04} \right]_{( ijkl )} $) it has only $15$ independent elements instead of $81$ in three dimensions.

A morphology metric suitable for the characterising systems of spherical particles should be rotationally invariant since the physics do not a priori designate a preferred direction. Thus, the tensor $W_1^{04}$ should not be directly used. Instead, rotational invariants are constructed. This is done by borrowing ideas from the theory of the elastic stiffness tensor.

The tensor $W_1^{04} (K)$ is rewritten in the Mehrabadi supermatrix notation \cite{MEHRABADI} as a $6\times6$ matrix:
\begin{equation}
\resizebox{.88\hsize}{!}{$M=\begin{bmatrix} 
S_{xxxx} & S_{xxyy} & S_{xxzz} & \sqrt{2} \, S_{xxyz} & \sqrt{2} \,S_{xxxz} & \sqrt{2} \, S_{xxxy}\\
S_{xxyy} & S_{yyyy} & S_{yyzz} & \sqrt{2} \, S_{yyyz} & \sqrt{2} \, S_{yyxz} & \sqrt{2} \, S_{yyxy}\\
S_{xxzz} & S_{yyzz} & S_{zzzz} & \sqrt{2} \, S_{zzyz} & \sqrt{2} \, S_{zzxz} & \sqrt{2} \, S_{zzxy}\\
\sqrt{2} \, S_{xxyz} & \sqrt{2} \, S_{yyyz} & \sqrt{2} \, S_{zzyz} & 2 \, S_{yzyz} & 2 \, S_{yzxz} & 2 \, S_{yzxy}\\
\sqrt{2} \, S_{xxxz} & \sqrt{2} \, S_{yyxz} & \sqrt{2} \, S_{zzxz} & 2 \, S_{yzxz} & 2 \, S_{xzxz} & 2 \, S_{xzxy}\\
\sqrt{2} \, S_{xxxy} & \sqrt{2} \, S_{yyxy} & \sqrt{2} \, S_{zzxy} & 2 \, S_{yzxy} & 2 \, S_{xyxz} & 2 \, S_{xyxy}\\ \end{bmatrix}$}
\end{equation}
where $S=W_1^{04} (K) /W_1 (K)$.

Then the six-tuple formed by the eigenvalues $\zeta_i$ of $M$ (in descending order) may be considered a symmetry fingerprint of the polyhedron $K$. It is invariant under rotation, scaling and translation of the polyhedron $K$.
Using the signature eigenvalue tupel $\zeta_i$  of $M$ it is possible to define a distance measure on the space of bodies induced by the Euclidean distance:
\begin{equation}
\Delta (K_1, K_2): =  \left( \sum_{i=1}^6 \left( \zeta_i(K_1) - \zeta_i(K_2)  \right)^{2}  \right)^{1/2}.
\label{eq-delta}
\end{equation}
$\Delta (K_1, K_2)$ is a pseudometric. It is positive definite, symmetric, the triangle inequality holds, however, the coincidence axiom $ \Delta(K_1,K_2) = 0 \Leftarrow K_1=K_2$ is only a implication and not an equivalence. For example $ \Delta (sphere, dodecahedron) = 0$. To distinguish dodecahedra from spheres one needs to employ even higher rank tensors. 

\begin{figure}[!tbp]
	\captionsetup[subfigure]{position=top,singlelinecheck=off,labelfont=bf,textfont=normalfont, justification=raggedright, margin=10pt,captionskip=-1pt }
	\hspace*{-0.1cm} 
	\subfloat[]{%
	\includegraphics[width=0.24\textwidth]{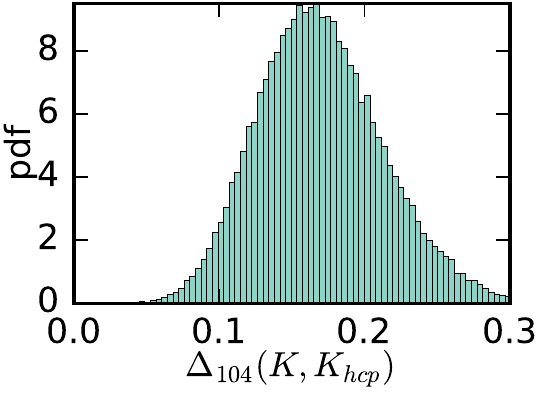}}
	\hspace*{-0.1cm} 
	\subfloat[]{%
	\includegraphics[width=0.24\textwidth]{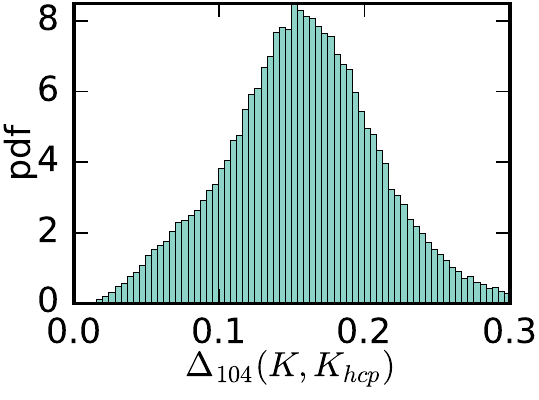}}
	\vspace{-0.5cm} 
	\hspace*{-0.1cm}
	\subfloat[]{%
	\includegraphics[width=0.24\textwidth]{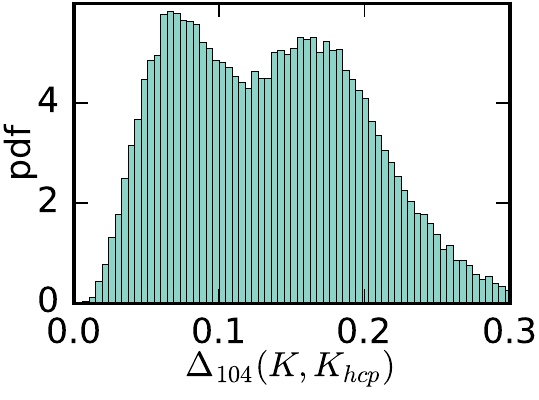}}
	\hspace*{-0.1cm} 
	\subfloat[]{%
	\includegraphics[width=0.245\textwidth]{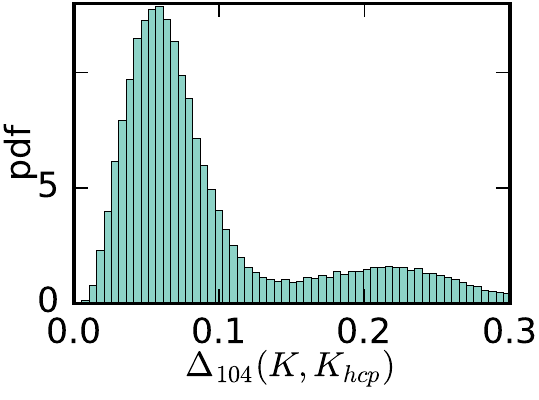}}
\caption{\label{fig-delta_hist} \small Calculating the minority phase (particle species 1) domain size via a Histogram method. The pseudometric $\Delta(K_{voronoi_cell},K_{hcp})$ (\ref{eq-delta}) of minority phase particles is plotted for increasing time from the top left to the bottom right panel. (a) $t=0.6 \, s$, (b) $t=0.6 \, s$, (c) $t=0.6 \, s$, (d) $t=0.6 \, s$. The separating parts can be interpreted as ordered domains (low $\Delta$ values) and disordered particles that not yet agglomerated (high $\Delta$ values) into a order domain. These plots are obtained for simulations with screening length ratio $\Lambda = 12$. For other values of the screening length ratio similar plots are obtained.}
\end{figure}

The MT 4 analysis carried out in this study is done in analogy to the MT 2 analysis. The symmetry metric $ \Delta(K_{voronoi \, cell},K_{hcp})$ (\ref{eq-delta}) is calculated locally for every Voronoi region of the minority phase (particle species 1) and for every time step in the simulation. Then a histogram is calculated for every time step. As illustrated in Fig. (\ref{fig-delta_hist}) these histograms are composed of two parts: The ordered, isotropic section (low $\Delta$ values) is separating from the disordered, anisotropic (small $\Delta$ values) section with increasing time steps. It is reasonable to assume that the isotropic Voronoi regions correspond to the homogeneous domains of agglomerating particles. To measure the size of these domains the number of Voronoi regions in the ordered, low $\Delta$ section of the histogram are counted. To that purpose, in order to distinguish ordered from disordered sections a cut off value ($\Delta_{thresh} = 0.12$) is chosen as the intersection point of the two sections at late time steps when demixing is well progressed. For simplicity this measure will be referred to as MT4 measure in the following.
\subsection{PS Method \label{subsec-ps}}
In the underlying previous study the onset and the first stages of the phase separation were characterised by the evolving domain size $L$, which was deduced from the time dependent average structure factor $S(k,t)$ \cite{dmx2_1367-2630-7-1-040, dmx4_0295-5075-93-5-55001,dmx20_Furukawa1984497} in a linear analysis. The position of the maximum of $S(k,t)$ was identified as $2 \pi /L $. The position of maximum at each simulation time step was determined by fitting the off-critical function $S(k,t) \propto \left( kL/2 \pi \right)^2 / \left[ 2+ \left( kL/2 \pi \right)^6 \right]$.

The results of this analysis are summarised in Fig. \ref{ps_vs_mt} (a) (triangles). See also supplemental movies S2 and S3 of \cite{demixing2010_PhysRevLett.105.045001}) In the SR dominated case $\Lambda=1$, when (\ref{eq-yukawa}) is reduced to the regular Yukawa form, the growth of domains of the minority phase is rather slow, and the evolution of $L(t)$ is characterised by relatively small growth exponents. In this case domains remain fuzzy at the simulation time scales and their shape is irregular. The increase of $\Lambda$, and therefore of the interaction range, sharpens the interfaces and makes the domains grow faster. As $\Lambda$ increases and LR interactions become more dominating they cause a large increase of the surface tension and faster demixing.

A more detailed description can be found in \cite{demixing2010_PhysRevLett.105.045001}.
\section{Results \label{sec-results}}
The results of the analysis with above explained methods and measures are presented in the following. To allow for comparison some plotted results were scaled: Each time series obtained by the PS measure and MT0 measure was scaled by multiplication with a factor such that the last (and also highest) point of these time series have the same value as the corresponding (same $\Lambda$) timeseries of the MT2 measure.
\subsection{Dynamic Range \label{subsec-dyn}}
\begin{figure}[!tbp]
%
%
	\hspace*{-0.35cm} 
	\includegraphics[width=0.48\textwidth]{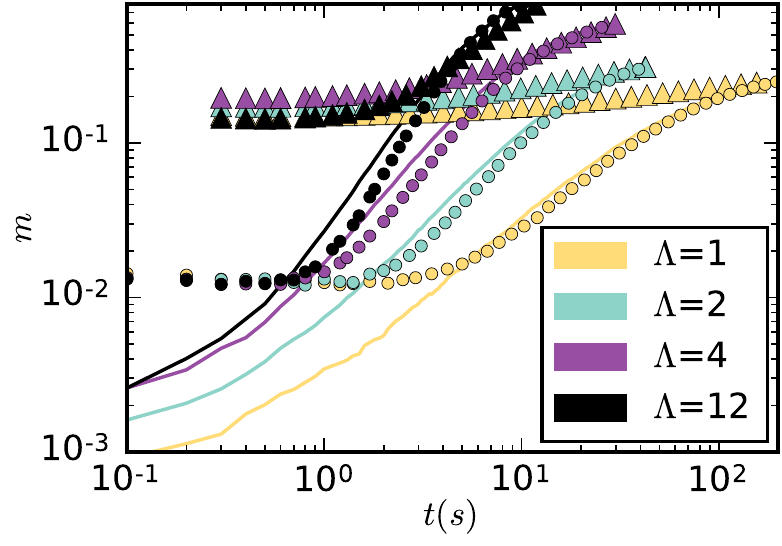}
\caption{\label{ps_vs_mt} \small Growth of the minority phase (particle species 1) domains. Different domain size measures $m \in \left\lbrace \text{PS}, \text{MT0}, \text{MT2} \right\rbrace $ are plotted as they change in time during the demixing simulation. (Details concerning measures and the simulation can be found in section \ref{sec1}.) \newline Triangles are obtained by the PS method, lines by the volume MT0 method, dots by the MT2 isotropy index method. Different colours indicate different screening length ratios $\Lambda$. For better visibility only a subset of all data points are displayed s.t. all neighbouring points have approximately equal spacing.}
\end{figure}	
Comparing the graphs of the domain size measure obtained by using the original power spectrum (PS) method (Fig. \ref{ps_vs_mt}, triangles) with the graphs of volume Minkowski functional (MT0 measure) analysis ( Fig. (\ref{ps_vs_mt} (a), lines)) we find that the volume functional yields a more sensitive measure for the growth of the minority phase (particle species 1) domains. 
Firstly one notices that both graphs reproduce a power law indicated by the affine graph in the double logarithmic plot. Secondly the significantly larger range of the volume curve and correspondingly its significantly steeper slope can be observed.

Furthermore the MT0 measure is able to resolve the domain growth right from the beginning of the simulation, whereas the PS measure does not respond until a certain threshold is reached. This might be due to the local nature of the MT0 method. Further investigation is necessary. Resolving domain growth right at the beginning reveals distinct phases (indicated by distinct slopes) of slower power law domain growth prior to the domain growth phase that was also found with the previous PS method.

Comparing the MT0 measure to the MT2 measure one can see that the capability of resolution in the beginning domain growth phases is lost. (Compare Fig. \ref{ps_vs_mt}.) However when the main domain growth phase is reached the MT2 measure becomes similar to the MT0 measure.

\begin{figure}[h!tbp]
%
%
	\hspace*{-0.35cm} 
	\includegraphics[width=0.48\textwidth]{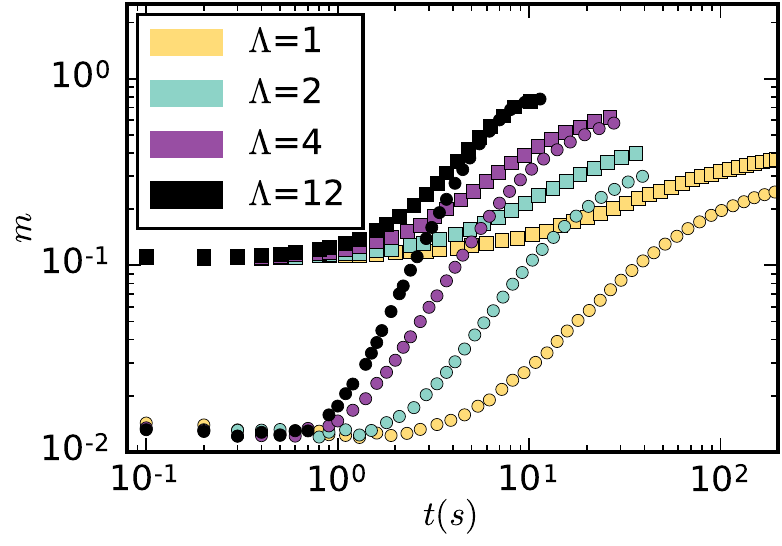}
\caption{\label{mt2_vs_mt4} \small Growth of the minority domains. Different domain size measures $m \in \left\lbrace MT2, MT4 \right\rbrace $ are plotted as they change in time during the demixing simulation. (Details concerning measure and the simulation can be found in section \ref{sec1}.) \newline Dots  are obtained by the $MT2$ isotropy index method, crosses by the MT4 symmetry metric method. Different colours indicate different screening length ratios $\Lambda$. For increasing values of $\Lambda$ (corresponding to slower demixing) a increasing difference can be observed in the maximal absolute level between the MT2 and MT4 measure. For better visibility only a subset of all data points are displayed as in Fig. \ref{ps_vs_mt}}
\end{figure}

Increasing the tensor rank of the Minkowski analysis and utilizing the MT4 measure we find similar results to the MT2 measure. The most significant difference is however the smaller dynamic range (and therefore also sensitivity) of the MT4 domain size measure. (Compare Figur \ref{mt2_vs_mt4}.) This might be due to the fact that only particle number or volume is required for any domain size analysis but the higher rank MT4 analysis also incorporates anisotropies that are not relevant in this regard. The Minkwski Tensor measures contain more information that is needed, when calculating back to domain size from higher rank tensors some of this information is lost. 

In conclusion, it was found that increasing the tensor rank of the Minkowski analysis from rank one to rank two and four results in ever decreasing dynamical range and therefore sensitivity. However even the rank 4 Minkowski tensor method performs better in terms of sensitivity than the previously used linear power spectrum methods.
\subsection{Difference in demixed domain size \label{subsec-diff}}
Comparing graphs of the MT2 and MT4 analysis we not only find differences in dynamical range and sensitivity as discussed in the previous chapter but also a difference in the absolute level these measures maximally yield. This might be another hint to the fact that the higher rank tensor analysis utilizes structural information beyond the mere volume information needed to calculate the domain size. However, it is more likely that this difference is just an artefact due to the simple method of separating the two histogram parts of the MT4 pseudometric $\Delta$ via a cut off value. Using Gaussian mixture methods might be preferable for future calculations with this measure. A cut off value approach was sufficient for the MT2 isotropy index $\beta$ since the ordered and disordered parts in the $\beta$ histogram are much further apart in the demixed state (compare Fig. \ref{fig-w120_hist}).

\subsection{Hints of universal behaviour \label{subsec-univ}}
\begin{figure}[!tbp]
	\captionsetup[subfigure]{position=top,singlelinecheck=off,labelfont=bf,textfont=normalfont, justification=raggedright, margin=10pt,captionskip=-1pt }
	\vspace*{-0.2cm} 
	\hspace*{-0.1cm} 
	\subfloat[]{%
	\includegraphics[width=0.255\textwidth]{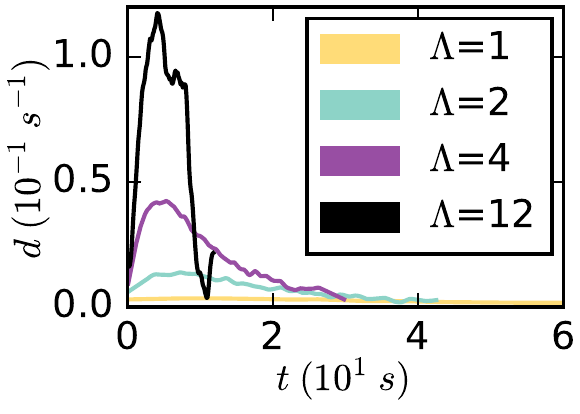}}
	\hspace*{-0.15cm} 
	\subfloat[]{%
	\includegraphics[width=0.235\textwidth]{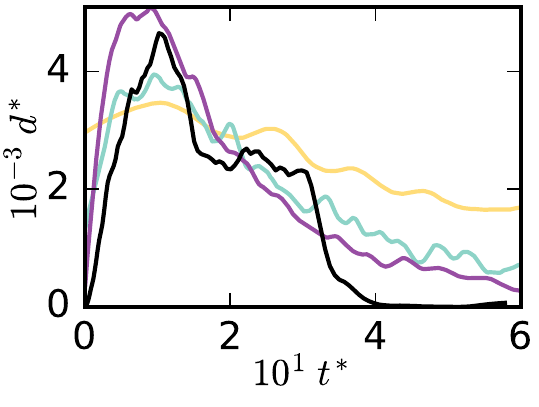}}
	\vspace{-0.5cm} 
	\hspace*{-0.1cm}
	\subfloat[]{%
	\includegraphics[width=0.255\textwidth]{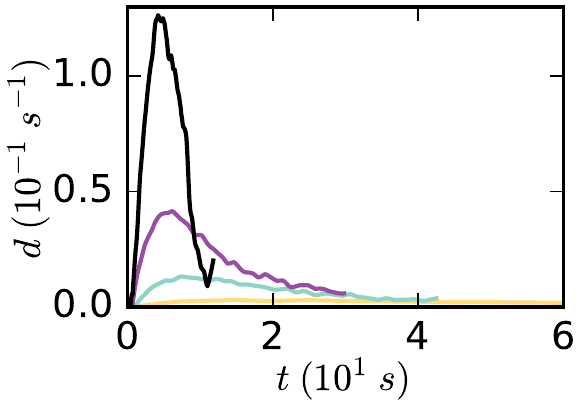}}
	\hspace*{-0.15cm} 
	\subfloat[]{%
	\includegraphics[width=0.235\textwidth]{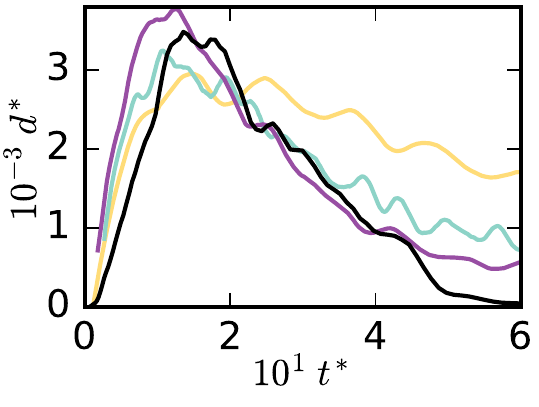}}
	\vspace{-0.5cm}
	\hspace*{-0.1cm} 
	\subfloat[]{%
	\includegraphics[width=0.255\textwidth]{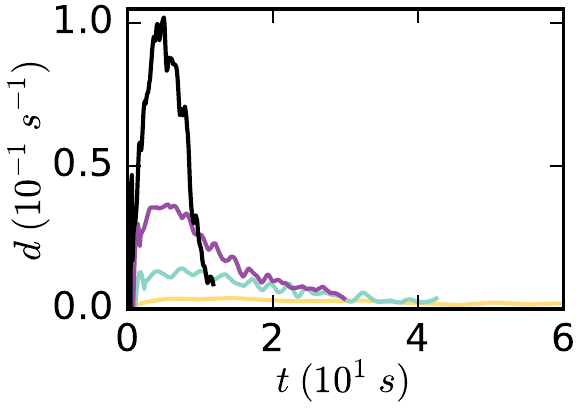}}
	\hspace*{-0.15cm} 
	\subfloat[]{%
	\includegraphics[width=0.235\textwidth]{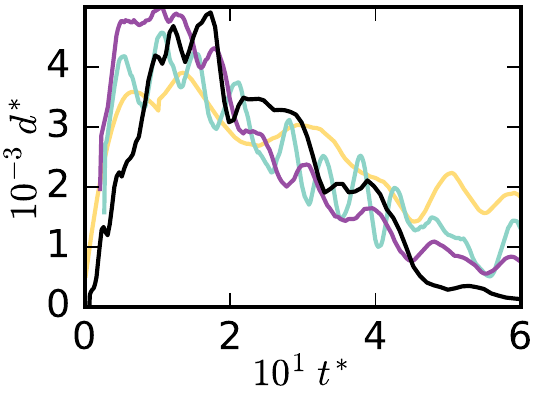}}
	\vspace{-0.5cm} 
	\hspace*{0.15cm} 
	\subfloat[]{%
	\includegraphics[width=0.24\textwidth]{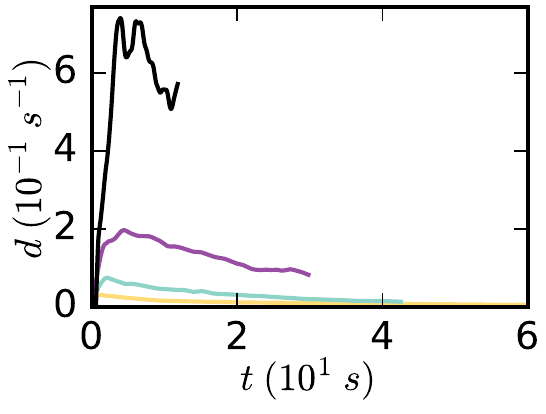}}
	\hspace*{-0.15cm} 
	\subfloat[]{%
	\includegraphics[width=0.235\textwidth]{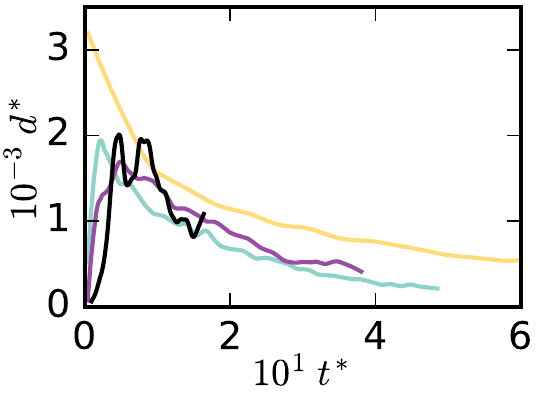}}%
\caption{\label{fig-dq} \small Left column: local gradients $d = \Delta m / \Delta t$ are plotted against time $t$. (a) $m=MT0$; (c) $m=MT2$; (e) $m=MT4$; (g) $m=PS$. Right column: Scaled local gradients $d^{*} =  ( d \times 1s )^{ \Lambda^{\mu}} \, , \, \mu \in \mathbb{R}$ are plotted against scaled time $t^{*} =  (t / 1 s )^{\Lambda^{\tau}} \, , \, \tau \in \mathbb{R}$. (b) $m=MT0$, $\mu=0.37$, $\tau=0.20$; (d) $m=MT2$, $\mu=0.405$, $\tau=0.24$; (f) $m=MT4$, $\mu=0.34$, $\tau=0.23$; (h) $m=PS$, $\mu=0.35$, $\tau=0.05$. The graphs are obtained by calculating the simple difference quotient $\Delta m / \Delta t$ for every point of the different MT and PS measures. Before and after differentiation the measure data was smoothed using a Savitzky-Golay filter \cite{savgol} of order $1$ and window sizes $w$ depending on the length of the time series ($\Lambda=1, w=7; \Lambda=2, w=17; \Lambda=4, w=23; \Lambda=12, w=101$). The legend for all sub Figures is identical to sub Fig. (a).}
\end{figure}	
\begin{figure}[!tbp]
	\captionsetup[subfigure]{position=top,singlelinecheck=off,labelfont=bf,textfont=normalfont, justification=raggedright, margin=10pt,captionskip=-1pt }
	\vspace*{-0.2cm} 
	\hspace*{-0.1cm} 
	\subfloat[]{%
	\includegraphics[width=0.245\textwidth]{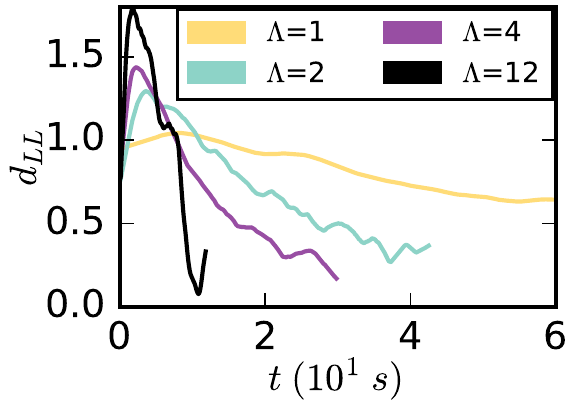}}
	\hspace*{-0.15cm} 
	\subfloat[]{%
	\includegraphics[width=0.24\textwidth]{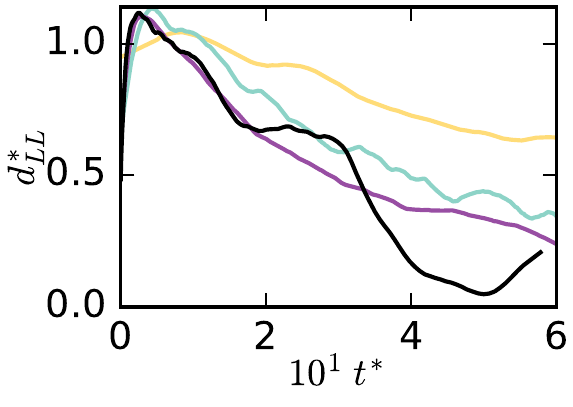}}
	\vspace{-0.5cm} 
	\hspace*{-0.1cm}
	\subfloat[]{%
	\includegraphics[width=0.245\textwidth]{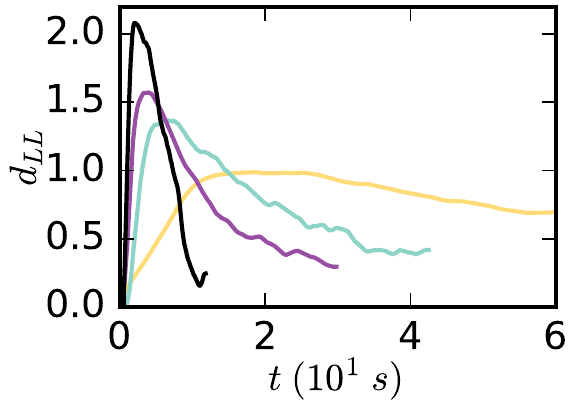}}%
	\hspace*{-0.15cm} 
	\subfloat[]{%
	\includegraphics[width=0.24\textwidth]{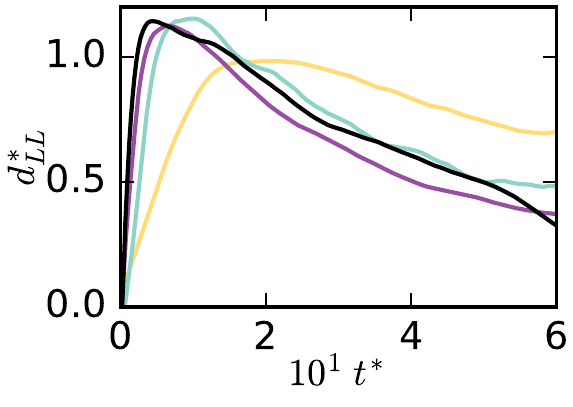}}
	\vspace{-0.5cm}
	\hspace*{-0.1cm} 
	\subfloat[]{%
	\includegraphics[width=0.245\textwidth]{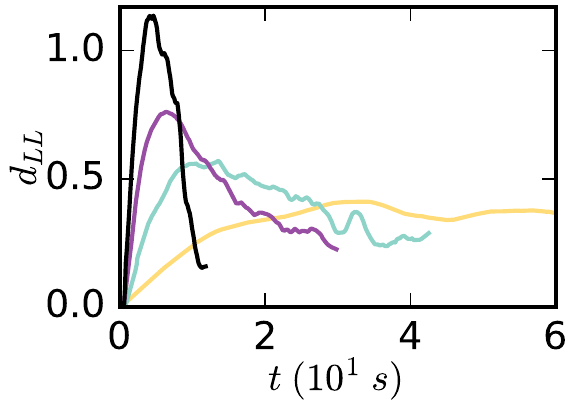}}
	\hspace*{-0.15cm} 
	\subfloat[]{%
	\includegraphics[width=0.24\textwidth]{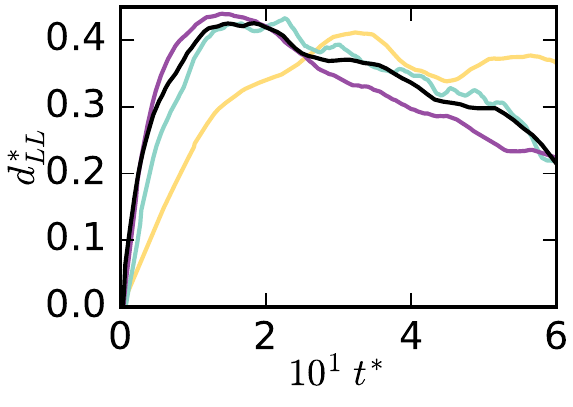}}
	\vspace{-0.5cm} 
	\hspace*{-0.1cm} 
	\subfloat[]{%
	\includegraphics[width=0.245\textwidth]{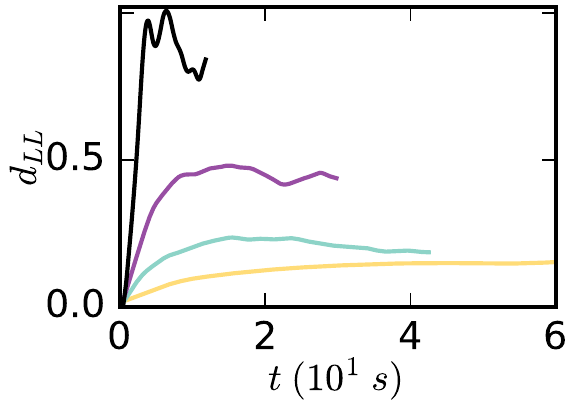}}
	\hspace*{-0.15cm} 
	\subfloat[]{%
	\includegraphics[width=0.245\textwidth]{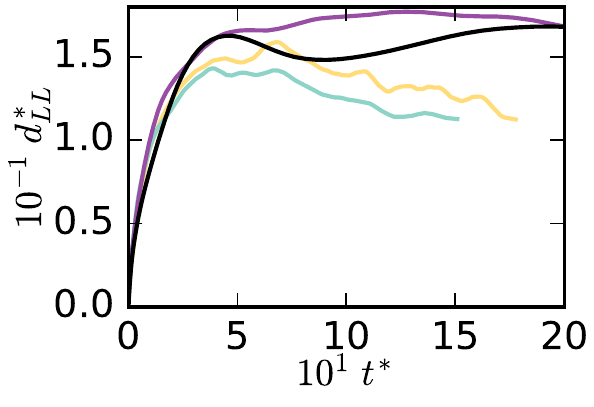}}
\caption{\label{fig-lldq} \small Left column: local power law exponents $d_{LL} = \Delta log(m \times 1s) / \Delta log(t/1s)$ are plotted against time $t$. (a) $m=MT0$; (c) $m=MT2$; (e) $m=MT4$; (g) $m=PS$. Right column: Scaled local gradients $d_{LL}^{*} =  d_{LL} / \Lambda^{\nu} \; , \; \nu \in \mathbb{R}$ are plotted against scaled time $t^{*} =  (t / 1 s )^{\Lambda^{\rho}} \, , \, \rho \in \mathbb{R}$. (b) $m=MT0$, $\nu=0.19$, $\rho=0.20$; (d) $m=MT2$, $\nu=0.24$, $\rho=0.26$; (f) $m=MT4$, $\nu=0.495$, $\rho=0.26$; (h) $m=PS$, $\nu=0.78$, $\rho=0.25$. The graphs for the different MT and PS measures are obtained by calculating $ (\Delta m \times t) / (\Delta t \times m) $ for every point which is equivalent to the simple difference quotient $\Delta log(m \times 1s) / \Delta log(t/1s)$. Smoothing was done as in Fig. \ref{fig-dq}. The legend for all sub Figures is identical to sub Fig. (a).}
\end{figure}
Qualitative differences in the demixing behaviour for different combinations of long and short-range interaction can be found by analysing the local gradients $d = \Delta m / \Delta t$ (Fig. \ref{fig-dq}) and local power law scaling behaviour $d_{LL} = \Delta log(m \times 1s) / \Delta log(t/1s) = (\Delta m \times t) / (\Delta t \times m)$ (Fig. \ref{fig-lldq}) of Minkowski measures $ m  \in \left\lbrace MT0, MT2, MT4 \right\rbrace $. In the unscaled plots (left columns) of Fig. \ref{fig-dq} and Fig. \ref{fig-lldq} a qualitatively different behaviour between the case $\Lambda = 1$ and the cases $\Lambda = 2,4,12$ can be observed: For $\Lambda = 2,4,12$ a significant increase and, after reaching a maximum, a significant decline in $d$ and $d_{LL}$ can be observed. However for $\Lambda = 1$ the local gradients $d$ and local power law exponents $d_{LL}$ do not show a significant decline after reaching their maximal level and remain almost constant.

This qualitative difference is also evident by considering scaled graphs of $d$ and $d_{LL}$ (right columns of Fig.  \ref{fig-dq} and Fig. \ref{fig-lldq} respectively). After finding a suitable scaling $\mu \in \mathbb{R}$ for $d$ (in the form $d^{*} =  ( d \times 1s )^{ \Lambda^{\mu}}$ ) and a time scale factor $\tau \in \mathbb{R}$ for $t$ ( $t^{*} =  (t / 1 s )^{\Lambda^{\tau}}$) the curves $d^{*}(t^{*})$ tend to collapse onto one curve for the cases with $ \Lambda \neq 1$. The scaled curves for $\Lambda = 1$ however do not follow the same behaviour: The scaled local gradient curves $d^{*}(t^{*})$ for $\Lambda = 1$ do not reach a maximal level comparable to the cases $\Lambda = 2,4,12$ and also decline much slower after reaching their maximal value. The same procedure is applicable to the local power law scaling exponent $d_{LL}$: By using the scaling $d_{LL}^{*} =  d_{LL} / \Lambda^{\nu} \; , \; \nu \in \mathbb{R}$ and $t^{*} =  (t / 1 s )^{\Lambda^{\rho}}\, , \, \rho \in \mathbb{R}$ the curves $d_{LL}^{*}(t^{*})$ tend to collapse onto one curve for cases $ \Lambda \neq 1$. The scaled local power law exponent curves $d^{*}_{LL}(t^{*})$ for $\Lambda = 1$ reach their maximal level at later times (for MT2 and MT4 analysis) compared to the cases $\Lambda = 2,4,12$ and also decline slower after reaching their maximal value.

This qualitatively different demixing behaviour can also be observed in the data of particle position of detected demixed domains when compared during common scaled times $t^*$ as in Fig. \ref{fig-particles} (or in the supplemental movies in the online version of this paper \cite{supp}). 

In the long range dominated cases ($\Lambda = 2,4,12$) demixing seems to occur in two stages: At first neighbouring particles agglomerate (scaled times $t^* \lesssim 10$, before gradient or power law exponent reaches maximal value ) then the agglomerated domains start to merge in cascades at scaled times $t^* \sim 10$ when the gradient or power law exponent of the Minkowski measures reach their maximal level. When the domains start to merge, after the agglomeration phase, the number of demixed particles is only growing slowly and correspondingly the graphs of gradient and power law exponent start to decay. In this cascade phase the main demixing mechanism is the merging of already demixed domains. The higher the screening length ratio is, the more cascades are happening and the larger are the resulting domains. For the screening length ratio $\Lambda = 1$, however, there is only agglomeration and no cascades are happening in which these agglomerated domains merge explaining the qualitatively different behaviour of gradient and power law exponent curves in Fig. \ref{fig-dq} and Fig. \ref{fig-lldq}. 
\begin{figure*}[!tbp]
	\captionsetup[subfigure]{position=top,singlelinecheck=off,labelfont=bf,textfont=normalfont, justification=raggedright, margin=10pt,captionskip=-1.2pt }
	\vspace{-0.5cm}
	\hspace*{-0.3cm} 
	\subfloat[]{%
	\includegraphics[width=0.25\textwidth]{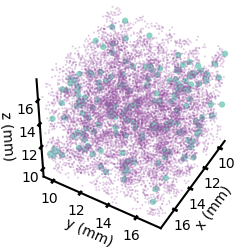}}
	\hspace*{-0.2cm} 
	\subfloat[]{%
	\includegraphics[width=0.25\textwidth]{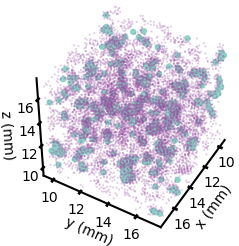}}
	\hspace*{-0.2cm}
	\subfloat[]{%
	\includegraphics[width=0.25\textwidth]{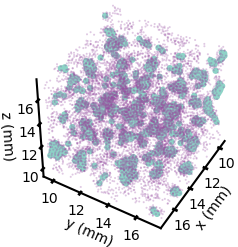}}%
	\hspace*{-0.2cm} 
	\subfloat[]{%
	\includegraphics[width=0.25\textwidth]{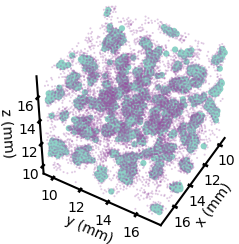}}
	\vspace{-0.7cm}
	\hspace*{-0.3cm} 
	\subfloat[]{%
	\includegraphics[width=0.25\textwidth]{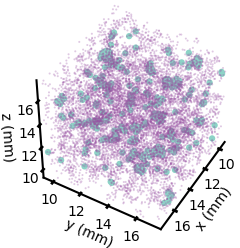}}
	\hspace*{-0.2cm} 
	\subfloat[]{%
	\includegraphics[width=0.25\textwidth]{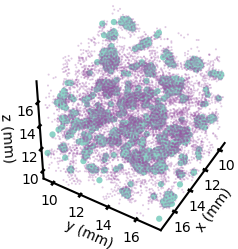}}
	\hspace*{-0.2cm} 
	\subfloat[]{%
	\includegraphics[width=0.25\textwidth]{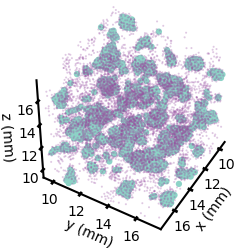}}
	\hspace*{-0.2cm} 
	\subfloat[]{%
	\includegraphics[width=0.25\textwidth]{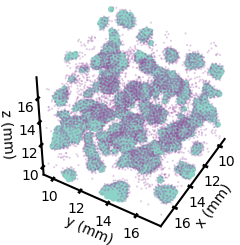}}
	\vspace{-0.7cm} 	
	\hspace*{-0.3cm} 
	\subfloat[]{%
	\includegraphics[width=0.25\textwidth]{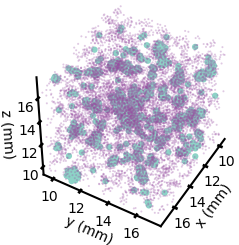}}
	\hspace*{-0.2cm} 
	\subfloat[]{%
	\includegraphics[width=0.25\textwidth]{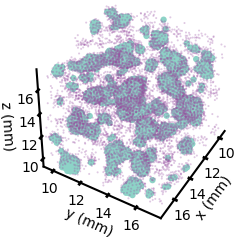}}
	\hspace*{-0.2cm} 
	\subfloat[]{%
	\includegraphics[width=0.25\textwidth]{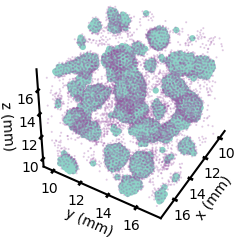}}
	\hspace*{-0.2cm} 
	\subfloat[]{%
	\includegraphics[width=0.25\textwidth]{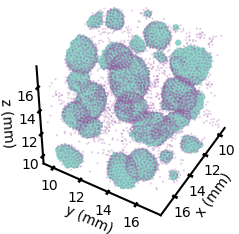}}
	\vspace{-0.7cm} 
	\hspace*{-0.3cm} 
	\subfloat[]{%
	\includegraphics[width=0.25\textwidth]{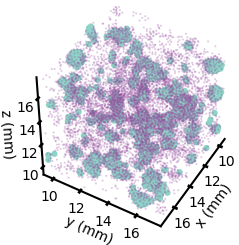}}
	\hspace*{-0.2cm} 
	\subfloat[]{%
	\includegraphics[width=0.25\textwidth]{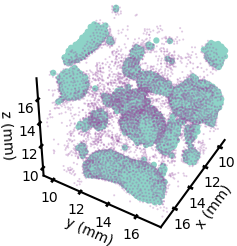}}
	\hspace*{-0.2cm} 
	\subfloat[]{%
	\includegraphics[width=0.25\textwidth]{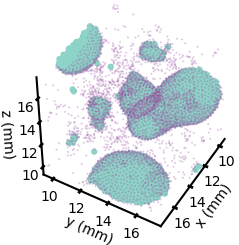}}
	\hspace*{-0.2cm} 
	\subfloat[]{%
	\includegraphics[width=0.25\textwidth]{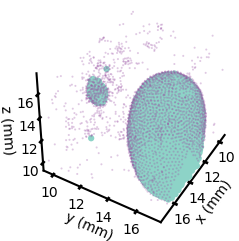}}
\captionsetup{width=2.0\textwidth}
\caption{\label{fig-particles} \small Particles of the minority phase (species 1) detected by the MT2 analysis as contributions to demixed domains are color coded in turquoise, the remaining species 1 particles are color coded in violet and are smaller in size. To assure visibility of the demixing process the particles of species 2 are not shown. They are homogeneously dispersed around the particles of species 1. The rows indicate different screening length ratios (The first row corresponds to $\Lambda =1$, the second row to $\Lambda =2$, the third row to $\Lambda =4$, and the fourth row to $\Lambda =12$). The first column corresponds to the scaled time $t^* = 5$ (corresponding to the agglomeration phase), the second column to $t^* = 15$ (agglomeration phase finished for $\Lambda =2,4,12$ and first merging cascade processes begin), the third column to $t^* = 25$ (early merging cascades for $\Lambda =2,4,12$ finished) and the fourth column to $t^* = 50$. (end stages of merging cascades for for $\Lambda =2,4,12$). The case $\Lambda=1$ is qualitatively different: Only agglomeration but no merging cascades of demixed domains can be observed. The scaled times given here are mean values of the different scaling times provided in Fig. \ref{fig-dq} and Fig. \ref{fig-lldq}. These illustrations can be viewed as video files in the supplemental material of the online version of this paper \cite{supp}.}
\end{figure*}	
\begin{figure}[!tbp]
	\captionsetup[subfigure]{position=top,singlelinecheck=off,labelfont=bf,textfont=normalfont, justification=raggedright, margin=10pt,captionskip=-1pt }
	\hspace*{-0.0cm} 
	\subfloat[]{%
	\includegraphics[width=0.35\textwidth]{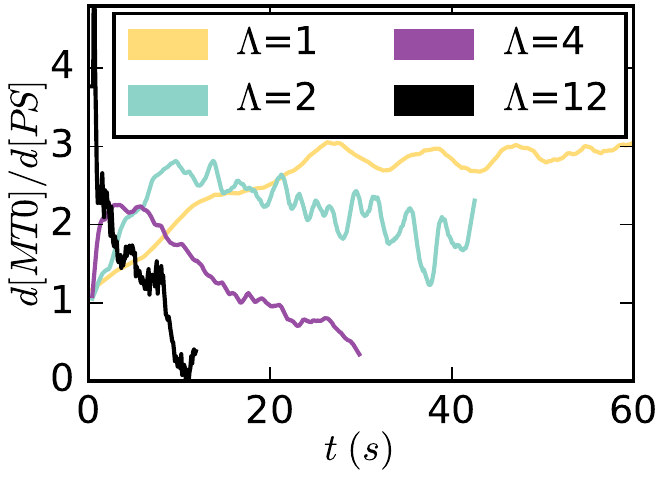}}
	
	\hspace*{-0.0cm} 
	\subfloat[]{%
	\includegraphics[width=0.35\textwidth]{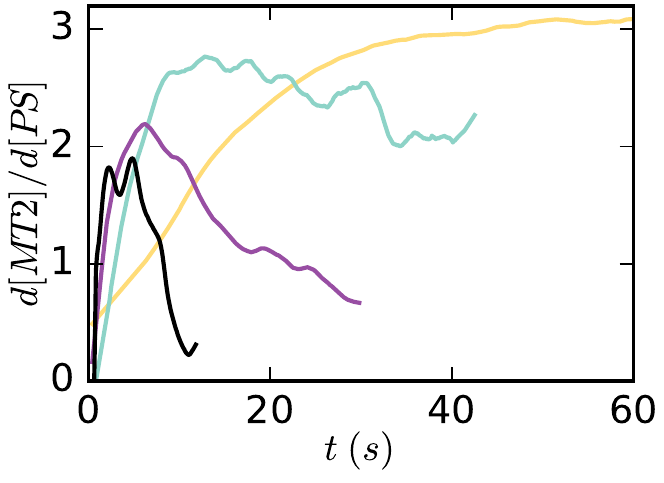}}
	
	\hspace*{-0.0cm}
	\subfloat[]{%
	\includegraphics[width=0.35\textwidth]{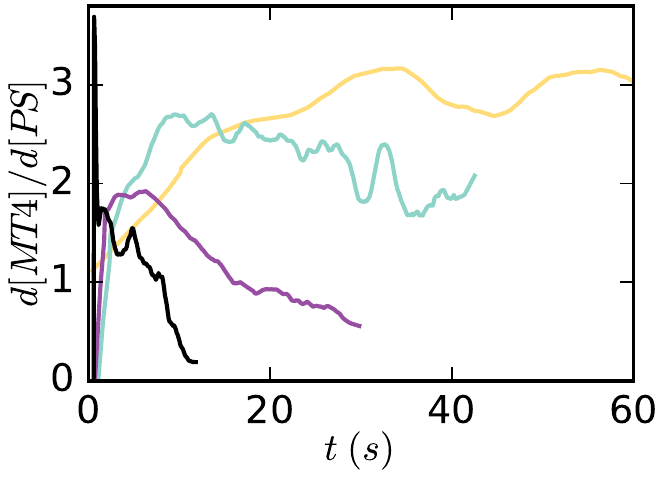}}%
\caption{\label{fig-dq_quot} \small Quotient of difference quotients $d$ of Minkowski measures $m$ and the PS measure. (a) $m=MT0$; (b) $m=MT2$; (c) $m=MT4$. Calculation and smoothing was done as in Fig. \ref{fig-dq}.}
\end{figure}
As illustrated in Fig. \ref{fig-dq_quot} the quotient of the difference quotients of the Minkowski measures and the PS measure is not constant and thus provides new information about the demixing behaviour. For $\Lambda > 1$ the graphs are rising until they reach a maximal value and then decline again. For the MT0 and MT4 measure there is an additional peak right at the beginning for $\Lambda =12$. Discarding this additional peak, the times when the maximal value is reached correspond approximately to the $t^*=15$, the time where the demixing changes from only agglomeration to the merging cascade dominated regime. This is plausible since the Minkowski measures are measuring the total volume of demixed regimes, whereas the PS measure measures the mean length scale of demixed regimes: When the merging cascade phase begins, the length scale of domains increases much faster compared to the agglomeration phase. However, the total volume still only increases due to residual agglomeration and does not change due to merging of already demixed domains.

For $\Lambda = 1$ the graphs are increasing in the beginning, however, they do not decline after reaching their maximal value but stay approximately constant again indicating a qualitatively different behaviour. Note that the quotient of local power law scaling exponents is reminiscent to the box counting dimension \citep{Mandelbrot1982Fractal} (also known as Minkowski–Bouligand dimension) as defined in fractal geometry due to the fact that it is equivalent to the quotient $log(V)/log(L)$ if self similar growth of domains is assumed. Since here $L$ is the mean domain size and not the box size, the quotient of local power law exponents is not the same as the box counting dimension.
\begin{figure}[!tbp]
	\captionsetup[subfigure]{position=top,singlelinecheck=off,labelfont=bf,textfont=normalfont, justification=raggedright, margin=10pt,captionskip=-1pt }
	\vspace*{0.1cm}
	\hspace*{-0.0cm} 
	\subfloat[]{%
	\includegraphics[width=0.35\textwidth]{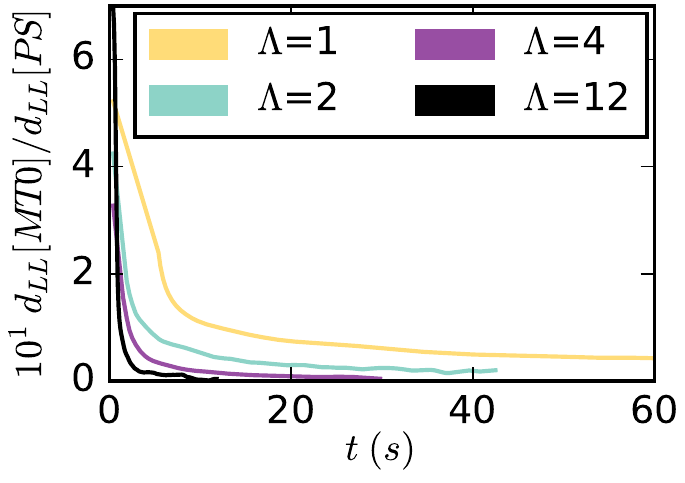}}
	
	\vspace*{0.05cm}
	\hspace*{-0.0cm}
	\subfloat[]{%
	\includegraphics[width=0.35\textwidth]{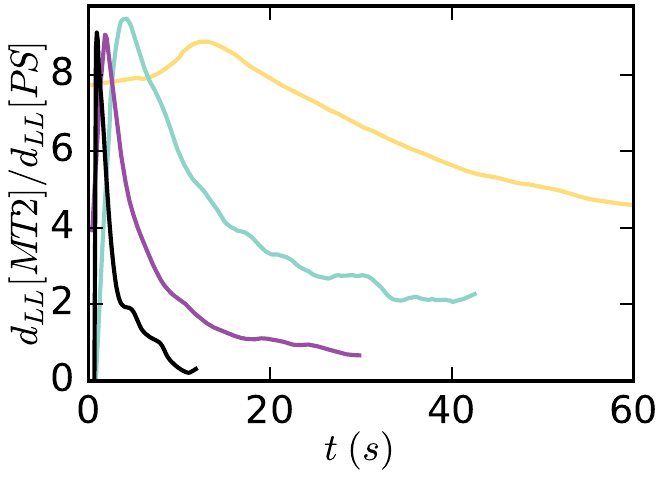}}
	
	\hspace*{-0.0cm}
	\subfloat[]{%
	\includegraphics[width=0.36\textwidth]{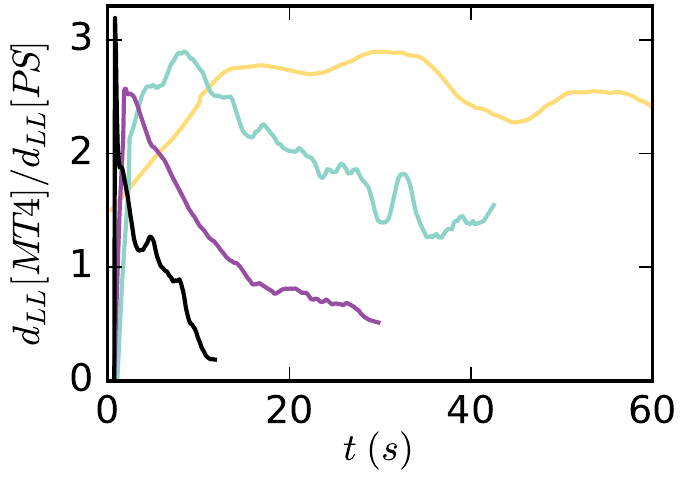}}%
\caption{\label{fig-lldq_quot} \small Quotient of local power law exponents $d_{LL}$ of Minkowski measures $m$ and the PS measure. (a) $m=MT0$; (b) $m=MT2$; (c) $m=MT4$. Calculation and smoothing was done as in Fig. \ref{fig-lldq}.}
\end{figure}
Similar behaviour to the behaviour of the quotient of gradients is found for the quotient of local power law exponents (Fig. \ref{fig-lldq_quot}). Additionally one can conclude that the MT0 measure has a much higher local power law exponent at the onset of demixing. This higher sensitivity will be advantageous for future studies of the onset of the demixing process.
\section{Conclusion and Outlook \label{sec-conclusion}}
Minkowski Tensor methods are a powerful tool for morphological characterisation and analysis. They can be superior to methods only utilizing the linear PS since they are inherently sensitive to non linear properties. They exhibit a higher sensitivity to changes in domain size and the measures exhibit a larger dynamical range. However utilizing the potential of higher rank tensors might not always prove to be worthwhile if not all the symmetry information they provide is needed. In this study employing rank 4 tensors proved to be unnecessary in order to obtain simple domain size information. 

Employing Minkowski Tensor methods reveals two qualitatively different demixing scenarios for simulations employing interactions with only one length scale ($\Lambda =1$) and different length scales ($\Lambda = 2,4,12$) which power spectrum analysis was not able to detect. Demixing occurs even for small screening length ratios $\Lambda=1$. However, it is not only slower compared to long range dominated ($\Lambda=2,4,12$) regimes but also shows qualitatively different behaviour: In the long range dominated cases ($\Lambda = 2,4,12$) demixing seems to occur in two stages: At first neighbouring particles agglomerate then the agglomerated domains start to merge in cascades. In this cascade phase the main demixing mechanism is merging of already demixed domains. The higher the screening length ratio is, the more cascades are happening and the larger are the resulting domains. For the screening length ratio $\Lambda = 1$ however there is only agglomeration and no cascades are happening in which these agglomerated domains merge. This is also evidenced by the universal behaviour of domain size measures based on Minkowski functionals and tensors. Further study of the universality is necessary. Future studies will try to concretise the hints of universal behaviour and qualitative differences for different screening length ratios. Questions like: How does the transition between qualitatively different demixing behaviour depend on other values for the screening length ratio and how this transition occurs, can be answered by means of more extensive simulations. More extensive simulations can also be used to investigate the effects of demixing in not perfectly monodisperse complex plasmas and how strongly demixing effects depend on the particle size distribution.

It might be beneficial to further investigate the onset of this process. There, the discreteness effects play an essential role and might require separate particle resolved studies focused on the comparison with the results of the coarse grained approach used in the simulations that provided data for this study \cite{demixing2010_PhysRevLett.105.045001}. As demonstrated in this study, morphological Minkowski Tensor analysis could be superior to PS methods in the investigation of the onset of demixing.\\
The (linear) scaling behaviour for the domain growth of a binary Lennard-Jones (LJ) liquid is well known \cite{dmx_lj_PhysRevE.77.011503}. To further test the efficiency of the Minkowski Tensor analysis methods and to shed new light on demixing processes in systems with LJ-like interactions we will apply MT analysis to these classical systems as well.\\
The demixing process of binary complex plasmas is a subject of current scientific interest: Only recently new experimental evidence gained via means of new visualisation techniques, based on the use of fluorescent dust particles, was published \cite{0exp_bin}. A clear trend towards phase separation even for smallest size (charge) disparities was observed. Further research in this direction can benefit from Minkowski tensor analysis as presented here since it is sensitive to nonlinear properties and capable of quickly revealing new aspects of interest in data. Minkowski tensor analysis is probing deeper than linear power spectrum analysis, however still providing easily interpretable results founded on a solid mathematical framework.
\section*{ACKNOWLEDGMENTS \label{sec:ack}}
We thank Adam Wysocki for providing the simulation data. The free software Qhull (http://www.qhull.org/) was used to compute the Voronoi tesselation. Qhull uses the Quickhull algorithm for computing the convex hull. Some Minkowski Tensors are calculated based on code of the free software package Karambola (provided at http://theorie1.physik.uni-erlangen.de/research/karambola/).
\section*{\label{sec:citeref}}

\end{document}